# Beam energy distribution influences on density modulation efficiency in seeded free-electron lasers

Guanglei Wang[1], Chao Feng[2], Haixiao Deng[2*], Weiqing Zhang[1], Guorong Wu[1], Dongxu Dai[1], Dong Wang[2], Zhentang Zhao[2] and Xueming Yang[1†]

[1]*State Key Laboratory of Molecular Reaction Dynamics, Dalian Institute of Chemical Physics, Chinese Academy of Sciences, Dalian 116023, P. R. China*

[2]*Shanghai Institute of Applied Physics, Chinese Academy of Sciences, Shanghai, 201800, P. R. China*

The beam energy spread at the entrance of undulator system is of paramount importance for efficient density modulation in high-gain seeded free-electron lasers (FELs). In this paper, the dependences of high harmonic micro-bunching in the high-gain harmonic generation (HGHG), echo-enabled harmonic generation (EEHG) and phase-merging enhanced harmonic generation (PEHG) schemes on the electron energy spread distribution are studied. Theoretical investigations and multi-dimensional numerical simulations are applied to the cases of uniform and saddle beam energy distributions and compared to a traditional Gaussian distribution. It shows that the uniform and saddle electron energy distributions significantly enhance the performance of HGHG-FELs, while they almost have no influence on EEHG and PEHG schemes. A numerical example demonstrates that, with about 84keV RMS uniform and/or saddle slice energy spread, the 30[th] harmonic radiation can be directly generated by a single-stage seeding scheme for a soft x-ray FEL facility.

*Corresponding authors: [*]denghaixiao@sinap.ac.cn; [†]xmyang@dicp.ac.cn

## I. INTRODUCTION

In recent years, enormous progresses have been achieved in the seeded free-electron lasers (FELs), which hold great potential to deliver high brilliance radiation pulses with excellent longitudinal coherence in the extreme ultraviolet and even x-ray regions. The first seeding scheme, i.e., high-gain harmonic generation (HGHG) has been fully demonstrated at BNL [1-4] and is currently used to deliver coherent extreme ultraviolet FEL pulses to users at FERMI [5]. For a long time, it is thought that the frequency multiplication factor of a single-stage HGHG is usually limited within ~10, due to the tradeoff between the energy modulation and the energy spread requirement for exponential amplification process of FEL. Therefore, a complicated multi-stage HGHG scheme [6-8] has been theoretically proposed and experimentally demonstrated for short wavelength production from a commercially available seed laser.

Meanwhile, alternative seeding concepts are under investigation to enhance the frequency up-conversion efficiency. The well-known echo-enabled harmonic generation (EEHG) scheme with dual modulator-chicane system has the potential to work efficiently at several tens of harmonic number in a single-stage configuration [9-10]. Recent efforts have demonstrated that the 3[rd], 4[th], 7[th] and 15[th] harmonics of the seed laser could be generated by a single stage EEHG [11-14]. More recently, another advanced seeding scheme termed phase-merging enhanced harmonic generation (PEHG) has been proposed [15-17], which benefits from the transverse-longitudinal coupling of the electron beam phase space. The novel design of PEHG equivalently suppresses the beam energy spread and future demonstrates remarkable harmonic up-conversion efficiency in a single-stage. Currently, a proof-of-principle PEHG experiment [18] is under way at Shanghai deep ultraviolet FEL test facility [19-21].

Up to now, the bunching performance assessment for seeded FELs is on the basis of assumption that the electron beam at the entrance of undulator has an energy spread of Gaussian distribution, which however is not true, e. g., in the specific case with a laser heater in the linac [22, 23]. Laser heater is widely utilized in high-gain FEL facilities to suppress the gain of the micro-bunching instability via Landau damping by controllable increasing the beam energy spread. It is found that a non-Gaussian energy distribution can be induced by a laser heater and inherited in the main LINAC section, depending upon details of the transverse overlap between the laser beam and the electron beam in the laser heater system. A recent experiment at FERMI [5] demonstrates that the non-Gaussian beam energy spread induced by the laser heater may expand the harmonic number of a single-stage HGHG to several tens [24-25].

Considering that the initial energy distribution of electron beam is one of the most critical elements in the bunching process of seeded FELs, in this paper, the possible beam energy distribution influences on density modulation efficiency in various seeded FEL schemes have been studied. A set of parameters of Shanghai soft x-ray free-electron laser facility (SXFEL) [26] is used in the discussions hereafter.

SXFEL aims at generating coherent 8.8 nm FEL pulses from 264 nm seed laser through a two-stage HGHG. In SXFEL, electrons from the photo-injector are firstly accelerated up to 130 MeV, and then sent into the laser heater system where the sliced beam energy spread is increased from 2keV to 8.4keV by a 792 nm Ti-sapphire laser. The pulse length of the laser beam is about 10 ps and the laser power density is controllable by adjusting the laser beam size and/or pulse energy to optimize the heating effect. After the laser heater, the electron beam will be further boosted to 840 MeV, and longitudinally compressed by a factor of 10. An electron beam with the normalized emittance of 1.0 μm-rad, sliced energy spread of 84 keV, bunch charge of 500 pC, and peak current of 600 A is expected at the exit of the LINAC for efficient FEL lasing.

As the sliced beam energy distributions summarized in Fig. 1, uniform and saddle distributions are investigated here in the density modulation process and compared to the previous Gaussian distribution case, under the same RMS energy spread of 84keV. It indicates that the beam energy distribution is of great importance for HGHG scheme, the frequency up-conversion number of a single-stage HGHG can be improved to 30 or even higher with a uniform or saddle electron energy distribution. EEHG and PEHG strongly change and reset the longitudinal beam phase-space, their performances depend weakly on the initial beam energy distribution. Moreover, for typical parameters of SXFEL, single-stage HGHG is capable for generating narrow bandwidth and saturated soft X-ray pulses with the existing radiator system.

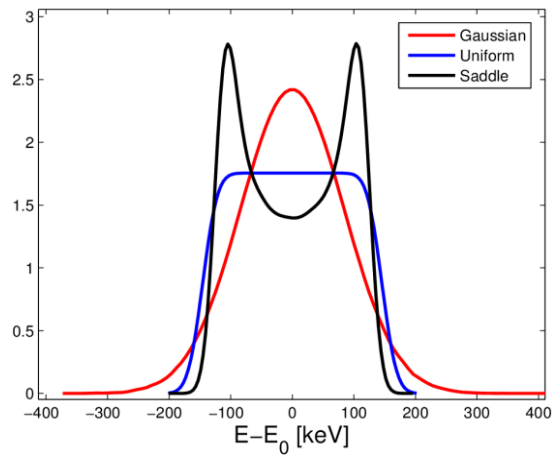

Fig. 1. The different beam energy distributions with same RMS deviation throughout this paper.

## II. EFFECTS OF ENENRGY DISTRIBUTION IN HGHG

Among the various seeding schemes, HGHG is the most compact and pioneering. The high harmonic bunching of HGHG can be described as [27]:

$$b = J_h(h\Delta\gamma_s D)\int dp f(p)e^{-ihD\sigma_E p}, \tag{1}$$

where $h$ is the harmonic number, $D = k_s R_{56}/\gamma$, $k_s$ is the wave number of the seed laser, $R_{56}$ is the strength of the dispersive chicane, $\gamma$ is the electron beam Lorentz factor, $\Delta\gamma_s$ is the seed laser-induced energy modulation amplitude and $J_h$ is the $h^{th}$ order Bessel function, $p = (E - E_0)/\sigma_E$ is the dimensionless energy deviation of a particle with an average energy $E_0$ and RMS energy spread $\sigma_E$. For a Gaussian energy distribution, following Eq. (1), the bunching factor can be written as,

$$b_G = J_h(hD\Delta\gamma_s)\exp(-\frac{h^2 D^2 \sigma_E^2}{2}). \tag{2}$$

For a saddle energy distribution, using the notations in ref. [24], i.e., the net longitudinal bunch length compression between the laser heater and the main undulator $C$, the energy modulation induced in the laser heater system $\Delta\gamma_h$ and the energy spread at the exit of the photo-injector $\sigma_H$, the bunching factor can be written as [22, 24],

$$b_S = \left|J_h(hD\Delta\gamma_s)\exp(-\frac{h^2 \sigma_H^2 C^2 D^2}{2}) \times J_0(hCD\Delta\gamma_h)\right|. \tag{3}$$

The predictions made by Eq. (2) and (3) have been analyzed intensively in Ref. [24]. The bunching factor draw back fast for a Gaussian energy distribution and this feature limits the feasibility of HGHG at high harmonics. For the saddle case, FERMI's experiment results show an FEL output pulse energy oscillation with the increase of the laser heating [24], which is a meaningful demonstration of Eq. (3).

If we assume a more ideal case that the electron energy is uniformly distributed between [$E_0$-τ/2, $E_0$+τ/2], the RMS energy spread is then changed to $\sigma_E = 0.5\tau/\sqrt{3}$. According to Eq. (1) and the law of Fourier transform for a rectangular pulse, the bunching factor for the uniform energy distribution at $h^{th}$ harmonic can be presented as

$$b_U = J_h(hD\Delta\gamma_s)\left|Sa(hD\tau/2)\right|. \tag{4}$$

To verify the abovementioned theoretical predictions and compare different cases, we carry out the single frequency simulations using the universal FEL simulating code GENESIS [28]. In these simulations, we take the main parameters of SXFEL as an example to illustrate the effects of different energy distribution on FEL density modulation process. Considering that the effective energy spread induced by the seed laser in HGHG is limited by the FEL parameter ρ for the requirement of exponential amplification in the final 8.8nm radiator, the energy modulation amplitude $A = \Delta E/\sigma_E$ is chosen to be about 5, and the optimal dispersive strength is chosen to be $k_1 R_{56} A \approx 1.2$ here.

Fig. 2 shows the bunching factor distributions at various harmonic numbers for different cases. The bunching factor oscillations are clearly seen for the uniform and saddle energy spread distribution cases. The amplitudes of the oscillations can be adjusted by setting the energy modulation amplitude and the strength of dispersive chicane. The simulation dots are all at the vicinity of the theoretical value, which is in good agreement with the derivation of Eqs. (2)-(4). This kind of bunching factor oscillation can be used to significantly extend the tuning range of the output wavelength of a single-stage HGHG down to very high harmonics, and makes the generation of soft X-ray FEL pulses in a single-stage HGHG possible.

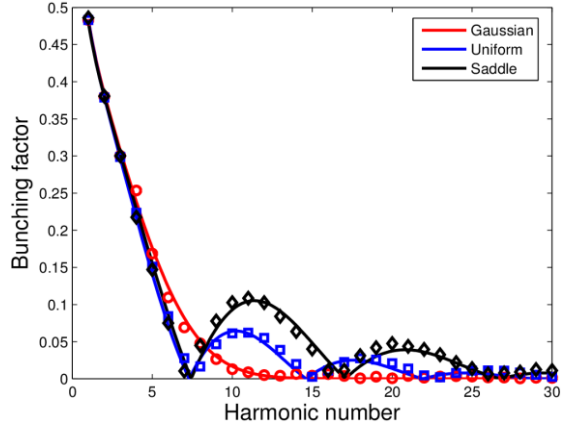

Fig. 2. The evolution of bunching factor with the harmonic number, the red circle is Gaussian results, blue square and the black diamond is for uniform and saddle respectively, the corresponding color line is theoretical deviation of Eqs. (2)-(4).

## III. EFFECTS OF ENENRGY DISTRIBUTION IN EEHG

In this section, we discuss the effects of energy distribution in EEHG scheme, following similar deviation in section II and the notations in ref. [10], assuming the frequency of the two seed laser in EEHG is identical, the EEHG bunching factor with typical Gaussian energy distribution can be represented as:

$$b_G = \left| \exp(-\frac{1}{2}(nB_1 + (m+n)B_2)^2) J_m(-(m+n)A_2B_2) J_n(-A_1[nB_1 + (m+n)B_2]) \right|. \quad (5)$$

$A_{1,2} = \Delta E_{1,2}/\sigma_E$ is the beam energy modulation amplitude induced in the modulators. For simplicity, we introduce the $B_{1,2} = R_{56}^{1,2} k\sigma_E/E$ for the strengths of the dispersive chicanes, $k$ represent the wave numbers of the seed laser, $m$ and $n$ are integers and the harmonic number of EEHG is $h = m + n$.

For the uniform energy distribution, the exponential term in Gaussian distribution is converted into a Sa function, and the EEHG bunching factor can be rewritten as:

$$b_U = \left| Sa(\tau(nB_1 + (m+n)nB_2/2)) J_m(-(m+n)A_2B_2) J_n(-A_1[nB_1 + (m+n)B_2]) \right|. \quad (6)$$

Following the derivation of Eq. (5), the bunching factor with a saddle energy distribution can be presented as:

$$b_S = \left| \exp(-\frac{1}{2}[(nD_1 + (m+n)D_2)C\sigma_H]^2) J_m(-(m+n)A_2B_2) J_n(-A_1[nB_1 + (m+n)B_2]) J_0(-C\Delta\gamma_h[(m+n)D_2 + nD_1]) \right| \quad (7)$$

According to the optimal condition of EEHG, the maximum bunching factor and the same sign of $B_{1,2}$ can be attained simultaneously when $n = -1$. Fig. 3 shows the simulation results of the bunching factor at different harmonic number. The EEHG parameters used here are $A_{1,2} = 2.5$ to obtain a same laser-induced energy spread with HGHG case mentioned above, the wavelength of the two seed lasers are 264 nm and each harmonic is optimized separately. The optimal relationship between the two dispersive chicanes for EEHG operation is:

$$B_2 = -\frac{n}{a}B_1 - \frac{\xi}{a}, \quad (8)$$

Where $\xi$ is the solution of $A_1[J_{n-1}(A_1\xi) - J_{n+1}(A_1\xi)] = 2\xi J_n(A_1\xi)$.

As mentioned above, the first strong chicane in EEHG will smear out the initial energy spread distribution of the electron beam, in Fig. 3, the EEHG bunching factor is nearly zero response to the energy distribution and the simulation results are in excellent agreement with the predictions of Eqs. (5)-(7).

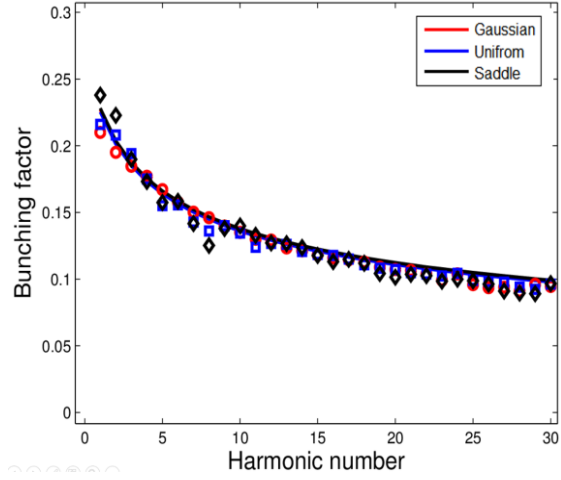

Fig. 3. The EEHG bunching factor vs. the harmonic number. The solid curves are calculated from Eqs. (5)-(7), the corresponding colorized dots represent the 3D simulation results.

## IV. EFFECTS OF ENENRGY DISTRIBUTION IN PEHG

PEHG scheme [15-17] combines the dogleg and the TGU modulator, which induce a transverse-longitudinal phase space coupling. When the transversely dispersed electrons pass through the TGU modulator, around the zero-crossing of the seed laser, the electrons with the same energy will merge into a same longitudinal phase. PEHG holds the great promise for generating fully coherent short-wavelength radiation. In the frame of 1D theory, the initial beam energy spread is artificially rearranged and fully suppressed in PEHG scheme by the so called phase-merging effect. Thus, the bunching factor of PEHG should be independent on the beam energy distributions.

If one takes the transverse effects into accounts, according to refs. [15-17], the PEHG bunching can be written as

$$b = J_h(hD\Delta\gamma_s)\exp(-\frac{h^2 D \sigma_x^2}{2\eta^2}) , \qquad (9)$$

where $\eta$ is the transverse dispersion of the dogleg and $\sigma_x$ is the transverse beam size. The reasonable beam size in the TGU modulator is $\sigma_x = \sqrt{\varepsilon_x L_m/2\gamma}$, with $\varepsilon_x$ and $L_m$ for the normalized horizontal emittance and the modulator length, respectively. It indicates that the bunching of PEHG is the same as the standard HGHG-FEL with an equivalent energy spread of $\sigma_{eff} = \sigma_x/\eta$. And PEHG bunching is immune to different energy distributions theoretically.

The PEHG bunching factor evolutions are illustrated in Fig. 4 for the three different energy distribution cases, where the energy modulation amplitude used in PEHG is also $A = 5$ to keep the same laser induced energy spread, the length of the modulator is $L_m \approx 1.2$, the dispersion of the dogleg is $\eta = 0.85$m and transverse gradient of TGU is $\alpha = 20\text{m}^{-1}$. The simulation dots with different energy distribution in PEHG have nearly the same tendency and the red line shows the prediction of Eq. (9) with an effective energy spread.

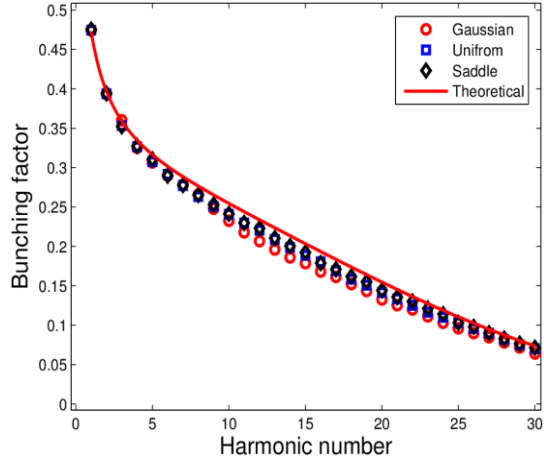

Fig. 4. The evolution of PEHG bunching factor .vs. the harmonic number, the dots are the simulation results, and the red line is theoretical deviation of the PEHG bunching factor with an effective energy spread.

Considering that the effective energy spread in PEHG comes from the transverse distribution of the electron beam, to be more comprehensive, different transverse beam distribution are also calculated, including Gaussian and uniform. The saddle case is ignored for realistic consideration. The bunching factor obtained for different transverse distributions are summarized in Fig. 5. The bunching factor in PEHG shows an oscillation for the uniform distribution as compared to the Gaussian. It is found that the bunching factor has an increase around the 50$^{th}$ harmonic for the uniform energy distribution, which is in good agreement with HGHG theory of Eq. (4) by using an energy spread of $\sigma_{eff}$. Moreover, the output peak power in tapered FEL can be dramatically improved for uniform transverse beam distribution case by enhancing the optical guiding in the undulator [29]. All these results indicate that the transverse shaping of the electron beam may be an effective way to improve the FEL performance.

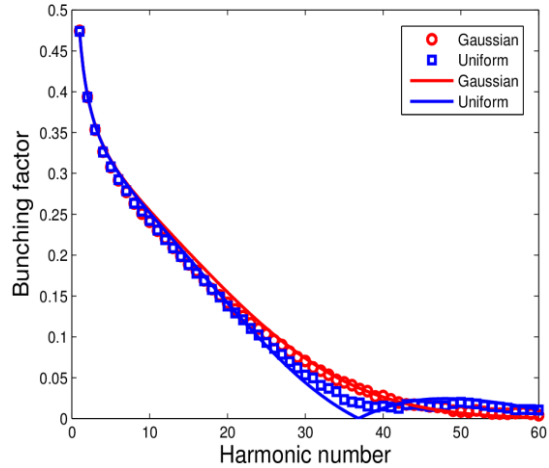

Fig. 5. The PEHG bunching factor vs. the harmonic number. The filled curves are calculated by Eq. (4) and Eq. (9), the colorized dots represent the simulation results.

## V. OPERATING SXFEL WITH SINGLE-STAGE HGHG

It has been widely discussed that, seeded FELs with total frequency up-conversion number of 30, e.g. SXFEL, can be realized by EEHG [30] and PEHG [15-17] in a single-stage configuration. In this section, we discuss the feasibility of operating SXFEL with a single-stage HGHG with initial saddle beam energy distribution as illustrated in Fig. 1. According to Eq. (3), the bunching factor of HGHG can be significantly enhanced with a saddle energy distribution. The optimized 30[th] harmonic bunching factor as a function of the HGHG scheme setup is shown in Fig. 6. One can find that the 30[th] harmonic bunching factor could be more than 4% for energy modulation amplitude $A$ around 6.5, which is strong enough for driving intense coherent radiation at the beginning of the radiator. Moreover, in view of the tradeoff between the seed laser induced energy spread and the available bunching factor, the energy modulation amplitude of $A = 6.5$ is moderate for FEL gain process in the radiator.

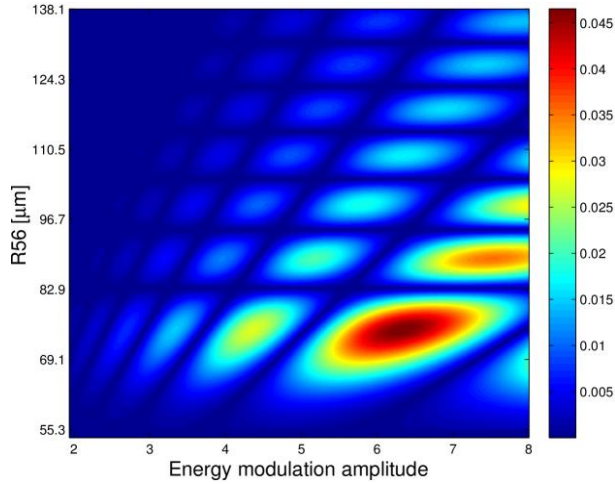

Fig. 6. The 30th harmonic bunching factor in a single-stage HGHG v.s. the energy modulation amplitude and the strength of the dispersion. A saddle sliced energy distribution with a RMS energy spread of 84keV is assumed.

With the above parameters, start-to-end tracking of the electron beam, including all the components of SXFEL, has been carried out. The electron beam dynamics in the photo-injector was simulated with ASTRA [31] to take into account space-charge effects. ELEGANT [32] was then used for the simulation in the remainder of the linac. Finally an artificial saddle energy distribution is loaded into GENESIS [28] at the entrance of the undulator system. In the simulation, the FWHM pulse duration of the 264nm seed laser is supposed to be 100fs. Fig. 7 shows the normalized output spectra and peak power evolution along the radiator. The relative FWHM bandwidth of 8.8nm FEL pulse is 0.1%. The noisy spike in FEL spectrum and 4 times larger bandwidth than the Fourier-transform-limited in the single-stage HGHG operation is induced by the shot noise, and mainly nonlinear beam energy chirp in the electron beam [33-36]. It is found that the peak power is approach to Gigawatt level, which is even larger than the original design of SXFEL with two-stage HGHG. The unsaturated peak power phenomenon in Fig. 7(b) is due to super radiant growth [37]. It is worth stressing that with the recent technology [38-39], the FEL performance can be future improved by removing the beam energy curvature [40].

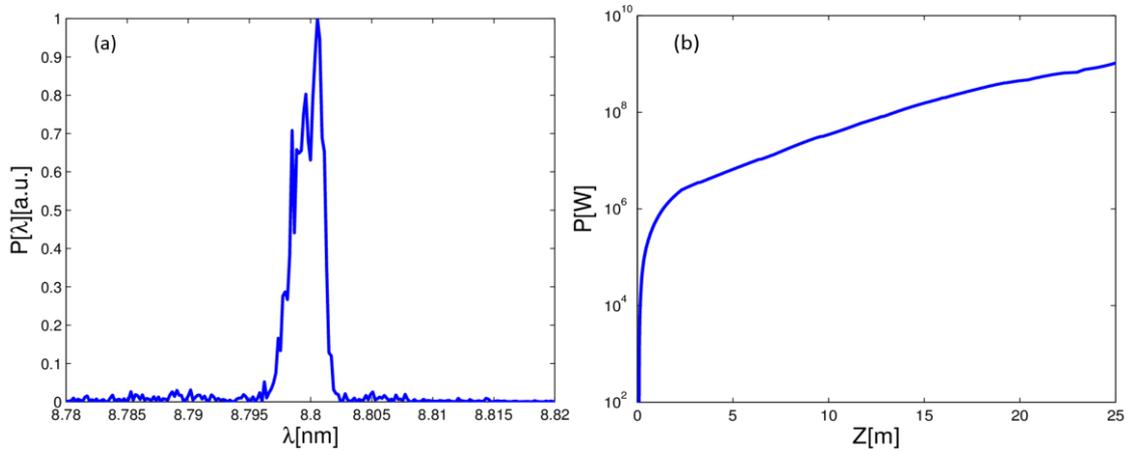

Fig. 7. The final 8.8nm radiation spectra (a) and peak power growth (b) of SXFEL with a single-stage HGHG.

## VI. CONCLUSION

In this paper, the sliced energy distribution effects on the bunching process in seeded FELs are investigated by using theoretical analysis and numerical simulations. It is found that bunching factor in different seeding mechanisms have quite different dependence on the beam energy distribution. EEHG and PEHG nearly have no response to different energy spread distributions, while a bunching factor oscillation happens in HGHG for uniform and saddle distributions. Moreover, such a bunching factor oscillation in HGHG can be adjusted by setting the energy modulation amplitude and the strength of the dispersive chicane, which indicates that the $30^{th}$ or even higher harmonic is possible in a single-stage HGHG with a moderate beam energy spread control by using the laser-heater system in the linac Thus, by manipulating the energy spread distribution, a single-stage HGHG may be used to cover much larger harmonic range than the theoretical predictions under the assumption of Gaussian beam energy spread distribution. However, in order to avoid the temporal coherence degradation due to the nonlinear beam energy curvature and coherent synchrotron radiation effect, a much shorter seed laser is preferred for high harmonic operation of single-stage HGHG. While in the advanced seeding configurations, i.e., EEHG and PEHG, a long seed lase pulse is allowed to entirely explore the ability of full electron bunch and thus enhance the average FEL brightness.

Finally, it is worth emphasizing that the different beam energy distributions were given artificially in our simplified analysis. And there are still rooms for further improvement in the subsequent studies, such as self-consistent start-to-end beam tracking, flexibility of beam energy spread control and accurate sliced beam energy spread measurement. It is likely that they will determine the frequency up-conversion limit achievable for different seeded FELs

# Acknowledgement

The authors are grateful to Bo Liu and Dao Xiang for helpful discussions. This work was supported by the National Natural Science Foundation of China (21127902, 11175240, 11205234 and 11322550).